\documentclass[usenatbib]{mn2e}
\usepackage{graphicx}
\usepackage{natbib}
\usepackage{amssymb, amsmath}

  \newcommand{\bc}{\begin{center}}
  \newcommand{\ec}{\end{center}}
  \newcommand{\be}{\begin{equation}}
  \newcommand{\ee}{\end{equation}}
  \newcommand{\Msun}{\>{\rm M_\odot}}
  \newcommand{\hMsun}{h^{-1}\>{\rm M_\odot}}
  \newcommand{\Mpc}{~h^{-1}~{\rm Mpc}}
  \newcommand{\Hunit}{~h ~{\rm km}~s^{-1}~{\rm Mpc}^{-1}}
  \newcommand{\Xunit}{\>h^{-2}{\rm erg}\,{\rm s}^{-1}}

  \newcommand{\rmag}{\>^{0.1}{\rm M}_r-5\log h}

\title[Fossil groups]
  {Properties of fossil groups in cosmological simulations and galaxy
    formation models}
\author[Weiguang Cui, et al.]
  {Weiguang Cui$^{1,2,5}$\thanks{wgcui@oats.inaf.it}, Volker~Springel$^{3,4,2}$,
Xiaohu Yang$^1$,
   Gabriella De Lucia$^6$, 
\newauthor and Stefano Borgani$^{5,6,7}$\vspace{0.2cm}\\ 
  $^1$Key   Laboratory	  for	 Research    in    Galaxies    and    Cosmology,
    Shanghai Astronomical Observatory; the Partner Group of MPA; \\  
    ~~Nandan Road 80,  Shanghai	200030,  China\\
  $^2$Max-Planck-Institut f\"{u}r Astrophysik, Karl-Schwarzschild-Stra\ss{}e  1,
    85740    Garching bei M\"{u}nchen, Germany\\
  $^3$Heidelberg Institute for Theoretical Studies, Schloss-Wolfsbrunnenweg 35,
69118 Heidelberg, Germany\\
  $^4$Zentrum f\"ur Astronomie der Universit\"at Heidelberg, Astronomisches
Recheninstitut,
    M\"{o}nchhofstr. 12-14, 69120 Heidelberg, Germany\\
  $^5$Astronomy Unit, Department of Physics, University of Trieste, via Tiepolo 11,
I-34131 Trieste, Italy\\
  $^6$INAF, Osservatorio Astronomico di Trieste, via Tiepolo 11, I-34131
Trieste, Italy\\
  $^7$INFN -- National Institute for Nuclear Physics, Trieste, Italy}

\begin{document}

\maketitle

\begin{abstract} 
  It has been a long-standing question whether fossil groups are just sampling
  the tail of the distribution of ordinary groups, or whether they are a
  physically distinct class of objects, characterized by an unusual and
  special formation history.  To study this question, we here investigate
  fossil groups identified in the hydrodynamical simulations of the GIMIC
  project, which consists of resimulations of five regions in the Millennium
  Simulation (MS) that are characterized by different large-scale densities,
  ranging from a deep void to a proto-cluster region.  For comparison, we also
  consider semi-analytic models built on top of the MS, as well as a
  conditional luminosity function approach.  We identify galaxies in the GIMIC
  simulations as groups of stars and use a spectral synthesis code to derive
  their optical properties.  The X-ray luminosity of the groups is estimated
  in terms of the thermal bremsstrahlung emission of the gas in the host
  halos, neglecting metallicity effects.  We focus on comparing the properties
  of fossil groups in the theoretical models and observational results,
  highlighting the differences between them, and trying to identify possible
  dependencies on environment for which our approach is particularly well
  set-up. We find that the optical fossil fraction in all of our
    theoretical models declines with increasing halo mass, and there is no
    clear environmental dependence.  Combining the optical and X-ray selection
    criteria for fossil groups, the halo mass dependence of the fossil groups
    seen in optical vanishes.  Over the GIMIC halo mass range we resolve best,
    $9.0 \times 10^{12} \sim 4.0 \times 10^{13} h^{-1}\Msun$, the central
    galaxies in the fossil groups show similar properties as those in ordinary
    groups, in terms of age, metallicity, color, concentration, and
    mass-to-light ratio. And finally, the satellite galaxy number distribution
    of fossil groups is consistent with that of non fossil groups. These
  results support an interpretation of fossil groups as transient phases in
  the evolution of ordinary galaxy groups rather than forming a physically
  distinct class of objects.
\end{abstract}


\section{Introduction} \label{sec:1}

Galaxy formation is one of the most complex processes in the evolution of the
Universe, and many aspects of it are still poorly understood.  Based on the
hierarchical formation hypothesis, galaxy clusters and groups form through the
assembly of many galaxies into a large common dark matter halo.  Information
about the merging history within a group is carried mainly by its primary
central galaxy and the satellite galaxies. If evolved in isolation for a long
time, the end-product of the merging in a group should be a system with a
central early-type galaxy, and few left-over satellites. Groups with a
strongly dominating central early-type galaxy, and a bright extended X-ray
halo with a cooling time scale longer than the group merging timescale, have
been called ``fossil groups'' -- because such an outcome could be most easily
understood if these are groups with a very early assembly, leaving enough time
for $L_\star$ satellite galaxies to merge away with the centre.

After being originally discovered by \cite{Ponman}, \cite{Jones2003} defined
fossil groups (FGs) by requiring that these groups need to have a difference
in absolute R-band magnitude of at least $\Delta M_{12} > 2$ mag between the
brightest and the second brightest galaxy located within half the projected
virial radius, and that their X-ray luminosity should exceed $L_{\rm X,bol}
\gtrsim 2.13\times 10^{42}h^{-2}{\rm erg~s}^{-1}$.  FGs have since been
analyzed in numerous observational studies
\cite[e.g.][]{Jones2003,Barbera,Santos,Mendes,Aguerri11}. However, due to a
rather low number of FGs identified observationally and the lack of X-ray
information in many cases, most works focused on the magnitude gap. This
magnitude gap has been studied in theoretical work through large sky surveys
\citep[e.g.][]{Bosch,Yang2008,M2006}, pure dark matter simulations
\citep[e.g.][]{Onghia2007,Beckmann}, semi-analytical models
\citep[e.g.][]{Sales,Diaz}, or through gas simulations \citep{Dariush2007}.
However, these studies do not agree well on the optical fossil fraction.

A common theme driving many of these investigations has been the desire to
understand whether fossil groups are a class of objects that are special in
their formation history \citep{Onghia2005,Dariush2007,Beckmann}, and whether
they would show different environmental dependences than ordinary groups
\citep{Beckmann,Diaz2010}. Some conflicting claims about this have been made,
and it is not yet clear whether fossil groups can be easily accommodated in
the leading $\Lambda$CDM model for cosmic structure formation.

In this paper, we study the environment of fossil groups in the { \it Galaxies
  Intergalactic Medium Interaction Calculation} (GIMIC) simulations, as well
as in two semi-analytic models (SAM) of galaxy formation, and one conditional
luminosity function (CLF) catalogue.  Furthermore, we also investigate the
properties of central galaxies in FGs and non-FGs through the GIMIC
simulations.  The GIMIC runs consist of re-simulations of five different
density regions selected from the volume of the Millennium Simulation (MS,
\citealt{SpringleNature}).  By identifying fossil groups in these regions, we
can study in detail how the large-scale environment impacts the properties of
these systems, and for the first time, we can carry out such an analysis in
full hydrodynamic simulations that start directly from cosmological initial
conditions. Furthermore, since the regions of GIMIC are drawn from the MS, we
can readily use the SAM catalogues of \citet[][hereafter D07]{Delucia}, and
\citet[][hereafter B06]{B06}, to carry out a region-by-region comparison.  In
addition, the galaxy catalogues based on the conditional luminosity function
model of \citet[][hereafter CLF]{Yang-prepare} have also been constructed for
the MS halos; we can hence select the same five regions here as well.  The
comparison of these catalogues for the same regions yields insightful tests of
the predictions of the theoretical models with respect to FG properties. We
also compare the predictions of these models with relevant observational data
to assess how closely the models reproduce the observed FG properties.

This paper is organized as follows. In Section~\ref{sec:2}, we introduce the
GIMIC simulations, and describe our methods to analyze the runs in terms of a
spectral synthesis code to compute galaxy properties and a simple approach to
calculate the X-ray luminosity of host halos. In Section~\ref{sec:3}, the SAM
and CLF catalogues we use are introduced. Section~\ref{sec:4} is devoted to a
description of the properties of the galaxy populations in the different model
catalogues. A comparison of the properties of FGs from these models is
presented in Section~\ref{sec:5}, along with their comparison with
observational data. Finally, we discuss our results and summarize our main
conclusions in Section 6.

\section{Analysis of the GIMIC simulations} \label{sec:2}

\subsection{Simulation set and galaxy identification}

In the ``Galaxies Intergalactic Medium Interaction Calculation'' (GIMIC)
project, five different regions of comoving radius $\sim 20\Mpc$ were selected
from the volume of the Millennium Simulation \citep{SpringleNature} and
re-simulated with hydrodynamics included.  The five regions have different
mean over-densities that deviate by $(-2, -1, +0, +1, +2)\, \sigma$ from the
cosmic mean, where $\sigma$ is the rms mass fluctuation on a scale of $\sim
20\Mpc$ at $z = 1.5$. The simulations included physics modules for gas cooling
and photoionization, quiescent star formation associated with supernovae
feedback, kinetic supernova feedback and chemodynamics, but no AGN feedback.
The GIMIC simulations were run with {\small GADGET-3}, an updated version of
the {\small GADGET-2} code \citep[last described in][]{Springel2005}, with the
physics implementation corresponding to one of the models studied in the
`Overwhelmingly Large Simulations' project \citep{Schaye2010}. The GIMIC
regions have been each evolved separately, and were run at several different
resolutions. The mass of the gas particles in the GIMIC simulations was
  $\sim10^7\hMsun$, sufficient to make the galaxy catalog complete down to a
  $r$-band absolute magnitude limit of about $-11.0$. Throughout this paper, we
adopt the same cosmological parameters as used in GIMIC and the parent MS
project: $\Omega_m = 0.25,$ $\Omega_{\Lambda} = 0.75,$ $\Omega_b = 0.045,$
$n_s = 1,$ $\sigma_8 = 0.9,$ $H_0 = 100 \Hunit$ with $h = 0.73$.  A detailed
description of further methodological aspects of the GIMIC simulations can be
found in \citet{Crain2009}.

Halos in the simulation were identified by the standard friends-of-friends
(FOF) algorithm, applied to the high-resolution dark matter particles with a
linking length of $b = 0.2$ in units of the mean inter-particle
separation. Baryonic particles were linked to groups by associating them with
the group containing the nearest high-resolution dark matter particle.
Substructures within FOF halos were then identified with the {\small SUBFIND}
algorithm \citep{Springel01,subfind_d}, which decomposes each FOF group into a
set of locally overdense and gravitationally bound structures.  For the larger
FOF halos of group- or cluster-size, this procedure reliably identifies the
different constituent galaxies that make up the groups.  These appear as
different subhalos inside the FOF groups, and can be directly compared to
observed galaxies.

\begin{figure*}
\includegraphics[width=1.0\textwidth]{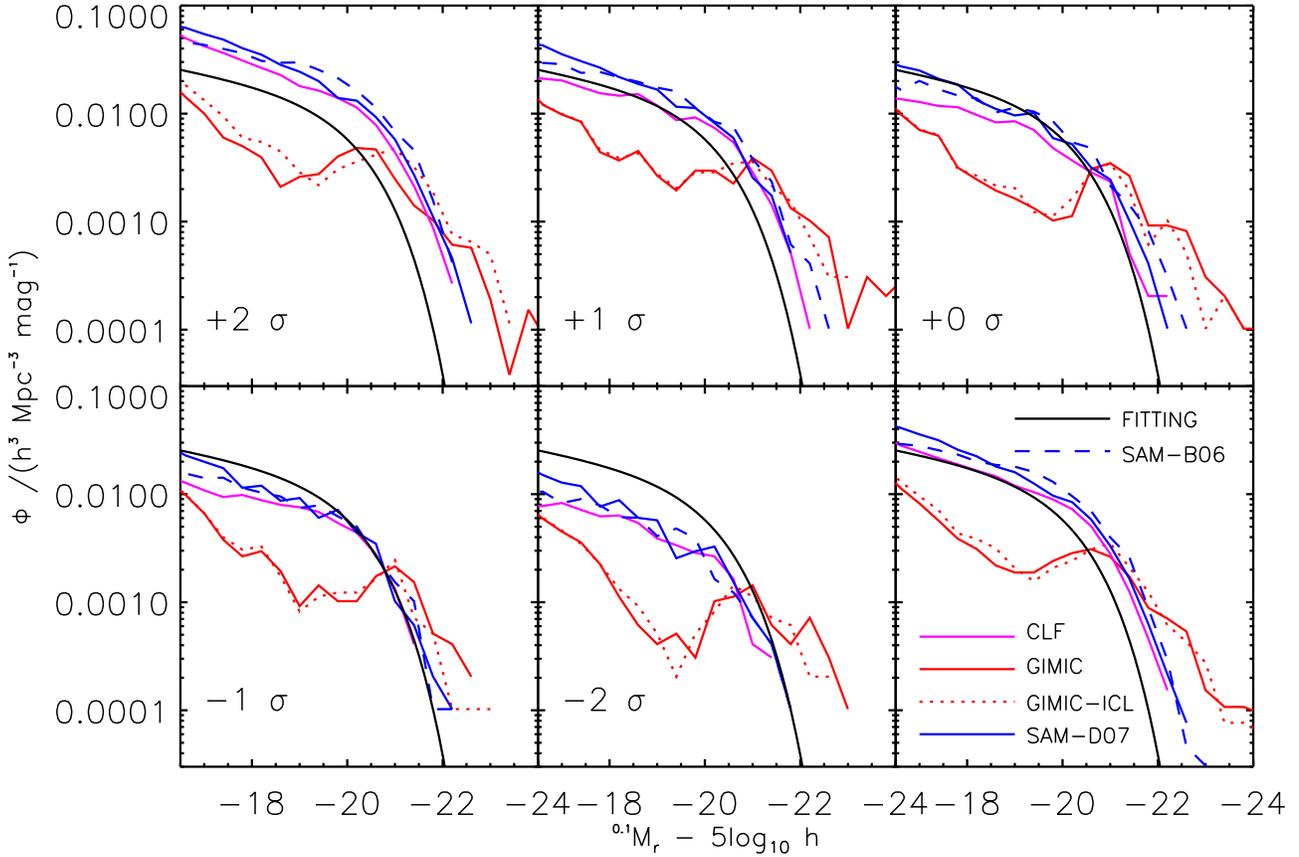}
\caption{ The galaxy luminosity functions of the GIMIC simulations (solid
    red line), GIMIC simulations without ICL (dotted red lines), the CLF
    catalogues (solid magenta) of \protect\cite{Yang-prepare}, and the two
    SAMs from B06 (dashed blue line) and D07 (solid magenta line) in the five
    different regions we analyzed, labeled as $+2\sigma$ to $-2\sigma$ from
    most overdense to most underdense. In all panels, the black solid line
  shows the Schechter form luminosity fitting function obtained by
  \citet{Blanton01} from the SDSS observational data.  The lower-right panel
  shows the averaged luminosity function of the five regions.}
\label{fig:LF}
\end{figure*}

In optical observations, it has been found recently that not all the emission
comes from stars that reside in the cluster's member galaxies. Instead, there
is also a smoothly distributed stellar component which is typically peaked
around the cluster's central galaxy, but extends to large radii. This diffuse
stellar component has been studied in a number of works. For instance,
\cite{Murante} and \cite{ICL} analysed high-resolution hydrodynamical
simulations of galaxy clusters, and pointed out that this intra-cluster light
(ICL) component accounts for up to $\sim 45\%$ of all the light in simulated
clusters and groups. This means we have to take this ICL component into
consideration and can not simply add it to the luminosity of the central group
galaxy, which would artificially boosts its luminosity. \citet{ICL} discussed
four different methods to identify the ICL component, all of them showing
quite similar results. So we here simply follow their simplest method,
limiting the aperture, to exclude the ICL of central galaxies. The cut radius
is given by \be r_{\rm cut} = 27.3\, h^{-1}{\rm kpc} \times \left(\frac{M_{\rm
    crit200}} {10^{15}\hMsun}\right)^{0.29}, \ee where $M_{\rm crit200}$ is
the mass within the radius enclosing a mean overdensity 200 times the critical
density of the Universe. In Figure~\ref{fig:LF}, we show the measured
luminosity function of galaxies in GIMIC, with (red solid line) and excluding
(red dotted line) the ICL stars. The two luminosity functions are not 
significantly different. Unless stated otherwise, we will always use central
galaxies without ICL component throughout this paper. 

\subsection{Optical galaxy properties}

We treat each stellar particle in the simulations as a simple stellar
population with the initial mass function (IMF) of \cite{Chab03}, with a mass
range of $0.1 - 100\Msun$, taking into account the age and metallicity of the
star particle.  This IMF matches the one that has also been used when running
the GIMIC simulations.  The spectrum of the stellar population is produced by
interpolating the SSP templates of \cite{BC03}.

The spectrum of each simulated galaxy is then simply obtained by adding up all
the stellar particles' spectra within the same subhalo: 
\be L_{\rm
  tot}(\lambda)=\sum_{i=1}^{N}{m_i \, L_{\rm SPP}(t_i,Z_i,\lambda)}, 
\ee
where $m_i$, $t_i$, and $Z_i$ are the mass, age, and metallicity of stellar
particle $i$, respectively, $N$ is the total number of stellar particles
within the subhalo, and $L_{\rm SPP}$ is the intrinsic SED per unit mass
interpolated from the SSP templates. Having obtained the spectra, we then
apply the SDSS filters in the AB system to get the galaxies' $u$, $g$, $r$,
$i$, and $z$-band magnitudes. We do not account for dust extinction in our
  procedure.

\subsection{X-ray luminosity of group halos}

Because one of the primary criteria often used to distinguish fossil groups
from ordinary groups is the X-ray luminosity, we measure this quantity
directly in the hydrodynamic simulations for all of our identified groups.
We adopt a simple approach to calculate the X-ray luminosities in the
  GIMIC simulations, where for simplicity we neglect metal lines and only
  consider the continuous X-ray emission that is due to thermal
  bremsstrahlung. Neglecting the contribution from metal lines can 
  provide an underestimate of X-ray luminosities, especially at relatively low
  X-ray temperatures, $T<3$ keV. This might undersestimate the bumber of X-ray
  selected fossil groups in our sample, but does not affect their fossil optical
  properties. We model the hot gas inside subhalos as only containing a
primordial mixture of hydrogen and helium, assuming that it is completely
ionized and optically thin.
 
The bolometric X-ray luminosity $L_X$ of a group halo can be obtained
by integrating the thermal bremsstrahlung emissivity over the halo's
volume. To calculate this integral, we can cast the volume integral
into a discrete sum over the SPH particles.  This gives \be
\begin{split}
L_X & = \frac{{\rm d}E}{{\rm d}t} = \kappa \int_V{\frac{0.72 g_B~
    T^{\frac{1}{2}}~ \rho^2}{m_p^2} {\rm d}V} \\ & = \frac{0.864\,
  \kappa}{m_p^2} \sum_{i = 1}^N {T_i^{\frac{1}{2}} \rho_i \,m_i},
\end{split}
\ee where $g_B$ is the bolometric Gaunt factor, which we set to 1.2,
$m_p$ the proton mass, $\kappa = (\frac{2\pi k T}{3
  m_e})^{\frac{1}{2}} \frac{2^5 \pi e^6}{3 h m_e c^3}$, with $m_e$ for
the electron mass, $e$ for the electron charge, and $T$ for the local
plasma temperature. In the above equation, the volume element
associated with each of the particles is $\Delta V_i = \rho_i / m_i$,
where $\rho_i$ and $m_i$ are the SPH density and particle mass for the
$i$-th particle, respectively.

\section{Semi-analytic and conditional luminosity function models} \label{sec:3}

\subsection{Semi-analytic galaxy catalogues}

The Millennium Simulation has been used extensively by the Munich and Durham
groups to construct so-called semi-analytic galaxy formation models in which
the evolution of galaxies is followed based on dark matter merger histories
measured from the simulation. The baryonic physics is described at a coarse
level with a simplified set of differential equations, allowing it to be
integrated forward in time comparatively quickly. The two semi-analytic galaxy
formation codes differ in their detailed assumption about the parameterization
of the relevant physics, making it interesting to also compare them with each
other, prompting us to consider both models.

Full details of the Munich semi-analytic model can be found in \cite{KH00},
\cite{Springel01}, \cite{DeLucia04}, and \cite{DeLucia06}, \cite{Croton06}. We
here analyze the model by \citet[][D07]{Delucia}, which is itself a modified
version of \cite{Croton06}, and includes a prescription for the growth and
activity of central black holes and their effect on suppressing the cooling
and star formation in massive halos.  D07 used the initial mass function of
\cite{Chab03}. Their publicly available catalogues include 26,787,155 galaxies
at $z=0$ in the whole Millennium volume.  The halo mass in this catalogue is
$M_{\rm crit200}$, which is defined as mass within the radius enclosing a mean
overdensity 200 times the critical density of the Universe. The dust
extinction is included in SDSS observer frame magnitudes for this galaxy
catalogue.

The Durham semi-analytic galaxy formation model is implemented in the {\small
  GALFORM} code \citep{Cole,Benson,Baugh,B06}. The semi-analytic galaxy
catalogue we analyze here is described in detail by \citet[][B06]{B06}, and
contains several enhancements compared to older {\small GALFORM} models, such
as the formation and growth of black holes, disc instabilities, improved
cooling calculation and AGN feedback. The B06 model adopts a Kennicutt IMF
with no correction for brown dwarf stars. The total number of galaxies number
in the final $z=0$ catalogue is 24,537,589. In this catalogue, dust extinction
is also included, but the magnitudes are in the Vega system, which we change
to the AB system to match the other catalogues used in this paper.

\subsection{Conditional luminosity function catalogues}

The conditional luminosity function (CLF) technique is a statistical procedure
to populate dark matter halos with galaxies such that the observed galaxy
luminosity function (or stellar mass function) and the clustering properties
of galaxies can be accurately reproduced.  The method has been described in
detail by \citet{Yang2003}, and has been refined over the years in numerous
studies \citep[][and references therein]{Yang-prepare}.

In a recent work, \citet{Yang-prepare} were able to take the full information
of subhalo merger histories into account when constructing the CLF as a
function of redshift.  Here we make use of the model parameters that are
constrained using the SDSS luminosity (and stellar mass) function and the
projected two point correlation functions.  MS dark matter halos are populated
with galaxies in two catalogues: one constructed with galaxy luminosities, the
other with galaxy stellar masses.

In the luminosity catalogue, each galaxy is assigned an $r$-band absolute
magnitude $\rmag$, which is $K+E$ corrected to redshift $z=0.1$
\citep{Blanton03}.  While in the stellar mass catalogue, the stellar mass is
assigned so that we can reproduce the stellar mass function obtained by
\citet{Yang2009}.

Note that as tested in \citet{Yang-prepare}, the CLF models and thus the
constructed galaxy catalogues can not only reproduce the observed SDSS
luminosity (stellar mass) functions, but also halo occupation numbers obtained
from the SDSS group catalogues. These catalogues contain roughly a total of
$10^7$ galaxies, slightly less than the two SAM models because of the brighter
luminosity (stellar mass) cut.

\begin{table*}
\begin{tabular}{|c|c|c|c|c|c|}
\hline
Model & $+2 \sigma$ & $+1 \sigma$ & $0 \sigma$ & $-1 \sigma$ & $-2 \sigma$ \\
\hline
GIMIC & 4,732 (2,583) & 1,418 (809) & 1,046 (651) & 898 (678) & 591 (471) \\
D07 & 14,516 (7,386) & 3,576 (2,226) & 2,394 (1,500) & 1,957 (1,525) & 1,261 (1,022) \\
B06 & 10,217 (5,298) & 2,431 (1,512) & 1,547 (961) & 1,212 (958) & 736 (620) \\
CLF & 13,553 (5,256) & 2,188 (1,610) & 1,437 (1,060) & 1,303 (1,115) & 822 (732) \\
\hline
\end{tabular}
\caption{The raw number of galaxies with $\rmag < -16$ in the five regions in
  each of the different theoretical galaxy catalogues. For reference, the
  second numbers given in brackets list the number of FOF-groups to which the
  galaxies belong. \label{table:galaxy_num} }
\end{table*}

\subsection{Matching of the GIMIC regions and comparing the four models}

The spatial location and radius of the five different density regions
re-simulated in GIMIC can be found in Table~1 of \cite{Crain2009}. It is hence
straightforward to cull the semi-analytic models and the CLF catalogues to the
same regions. Also, all the models have a parameter that identifies whether a
galaxy is the central galaxy in a group, or whether it is a satellite galaxy
within a group. We test only the central galaxies for membership in one of the
regions, and if this is the case, always use all the galaxies in the same
group, too.

The number of galaxies with $\rmag \leq -16$ and the number of FOF groups (in
brackets) in each region is listed in Table~\ref{table:galaxy_num}. The four
models do not agree with each other well in the number of galaxies and the
number of FOF groups in all five regions. The reason is that the four models
have different predictions for galaxies even in the same dark matter
halos. With the magnitude limit we applied, part of low mass galaxies and/or
groups are not included. It is interesting to see that the GIMIC model
produces fewer galaxies compared to other models in all regions for this
magnitude cut, especially in the $+2 \sigma$ region. From Figure~\ref{fig:LF},
the drop of $\Phi$ in GIMIC is mainly caused by the strong wind model.

Note that since the absolute magnitudes provided in SAMs and GIMIC are scaled
to redshift $z=0$, in order to properly compare with the SDSS observation, we
have converted them into the ones corresponding to redshift $z=0.1$ using the
average $K+E$ corrections provided by \citet{Blanton03,Blanton07}.  Throughout
the paper we use the same $\rmag$ notation. The four models also adopt
different definitions of halo masses.  For proper comparisons, we convert the
halo masses in the SAMs and GIMIC to the ones in the CLF model with a halo
abundance matching method. The final halo masses thus correspond to the FOF
halos whose mass function is well described by \citet{Sheth}. We apply
  this normalized halo mass of each model in the following sections.

\section{Basic Sample Properties} \label{sec:4}

Before we analyze the specific properties of fossil groups, we study here some
of the basic properties of our different model galaxy catalogues, and how they
compare to observational data.  We limit ourselves to a discussion of the
luminosity function, the properties of central galaxies in groups, and the
halo occupation distribution function.


\begin{figure*}
\bc
\includegraphics[width=1.0\textwidth]{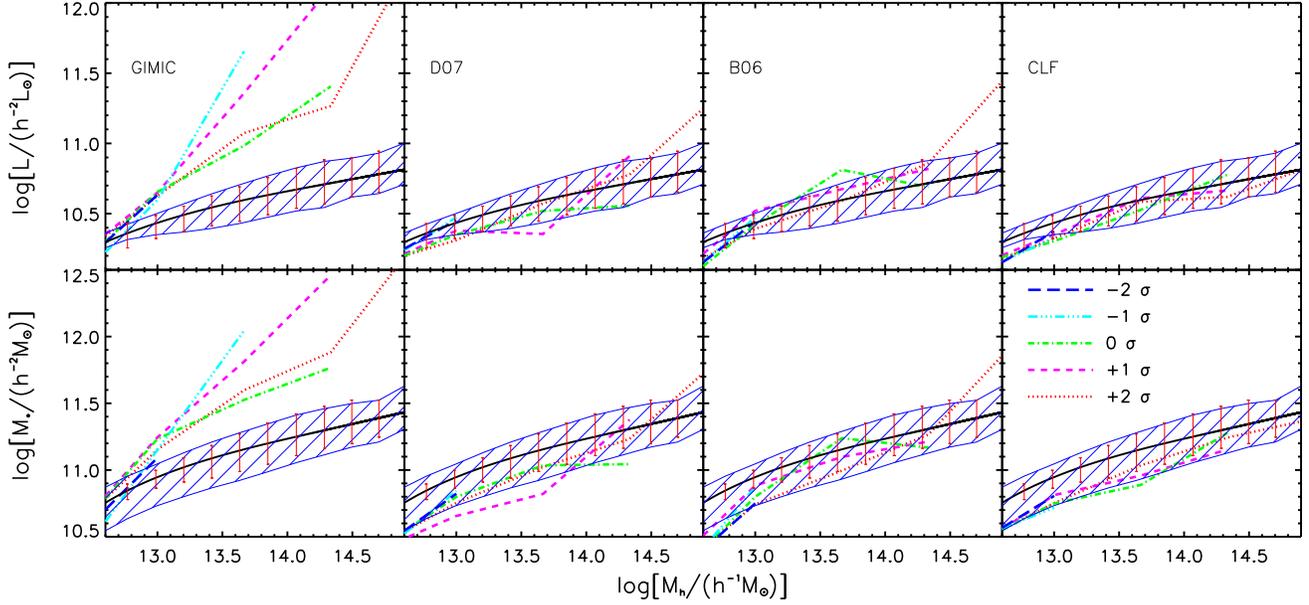}
\caption{{\em Upper panels}: Median r-band luminosity of the brightest cluster
  galaxy (BCG) as function of halo mass of four theoretical models, as
  labeled.  {\em Lower panels}: Median stellar mass of the most massive galaxy
  (MMG) as function of halo mass of the models. The solid lines in each panel
  give the results for the first ranked galaxies in the groups. The different
  colors and line styles encode the different density regions, according to
  the key shown in the lower-right panel.  For comparison, the solid black
  line shows the fitting formula of \citet{Yang2008} derived for the CLF
  model. The shaded area and error bars come from \citet{Yang2008}, and show
  the $68\%$ confidence regions of halo mass estimated through the group
  luminosity $M_L$ and group stellar mass $M_S$, respectively.
\label{fig:BCG} 
}
\ec
\end{figure*}

\begin{figure*}
\bc
\includegraphics[width=1.0\textwidth]{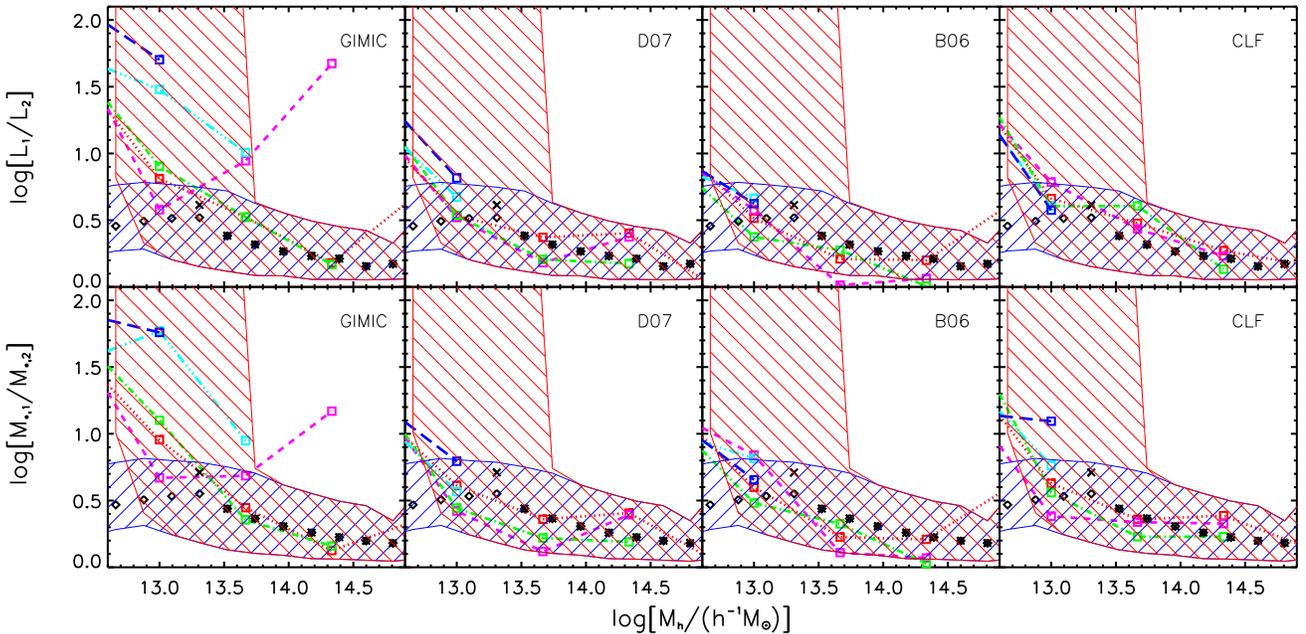}
\caption{ {\em Upper panels:} The luminosity ratio between first and second
  ranked galaxies, as a function of halo mass for our four theoretical models,
  as labeled.  {\em Lower panels:} The stellar mass ratio between first and
  second ranked galaxies as a function of halo mass for the same models.  The
  colored lines encode the different density regions, using the same key as in
  Figure~\ref{fig:BCG}.  The cross and square symbols represent an analysis of
  the observational data, only differing in the treatment of the second
  brightest galaxy \citep{Yang2008}.  The shaded areas come from
  \citet{Yang2008}, which indicate the corresponding $68\%$ confidence
  intervals for crosses (red shaded region) and squares (blue shaded region).
\label{fig:BCG_ratio}
}
\ec
\end{figure*}

\subsection{Luminosity function}

The luminosity function is one of the most fundamental characteristics of a
galaxy population. In Figure~\ref{fig:LF}, we show the GIMIC luminosity
function of the five different regions in separate panels, comparing with the
observational results of \cite{Blanton01} and with our semi-analytic and CLF
models. A Schechter function fit provides a good description of the
dust-corrected SDSS data over a range $-23 < \rmag < -16$, with best-fit
parameters $\phi_{\ast} = 1.46 \times 10^{-2}\,h^3\,{\rm Mpc}^{-3}$,
$M_{\star} = -20.83$, and $\alpha = -1.20$. The luminosity functions in the
different density regions are clearly different in their normalization. The
higher density regions have a higher number density of galaxies, as
expected. However, there is no obvious difference in the shape or
characteristic magnitude. In the lower-right panel of Figure~\ref{fig:LF}, the
total luminosity functions from the CLF and SAM models are also shown; they
exhibit a better match than GIMIC to the Schechter fit of \citet{Blanton01}
for the observational data. We note that the CLF and SAM luminosity functions
have a higher normalization than the observations. This is mainly due to the
fact that we have added all the galaxies in the five regions with equal
weight, for simplicity, while in reality there are fewer high density regions
than low density regions within a representative volume of the Universe.

It is obvious from Figure~\ref{fig:LF} that there is a substantial
disagreement between the shape of the LF of the GIMIC simulations and the
other theoretical models. In fact, the concave shape and bump-like feature in
the GIMIC luminosity function appears in all different density regions, and is
caused by the specific feedback model that was introduced in the hydrodynamic
simulations. In \cite{Crain2009} it was shown that the galaxy stellar mass
function has a similar trend \citep[see the right panel of figure $3$
  in][]{Crain2009}. They also investigated the origin of the disagreement in
some detail. At the lower mass end ($M_{\*} \lesssim 10^9 \hMsun$), the excess
number of galaxies reflects a reduction in the efficiency of SNe feedback in
poorly resolved galaxies, while at intermediate masses ($10^9 \sim 10^{10}
\hMsun$), the `dip' is caused by an efficiency peak in the feedback wind
model, that is even too effective in quenching star formation for galaxies in
this mass range.  Finally, at the high mass end ($M_{*} \gtrsim 10^{11}
\hMsun$), the overprediction of massive galaxies is mainly caused by gas
overcooling in the brightest cluster galaxies, which is not efficiently
counteracted by SN-driven feedback \citep[see also][]{Saro2006,Nuza}.
Resolving this problem may require the introduction of AGN feedback.
 
Taken at face value, the poorer agreement of the hydrodynamic GIMIC simulations
with the observational data may seem to suggest that these calculations are
inferior to the other theoretical models. It needs to be taken into account,
however, that the hydrodynamic simulations do not involve much tuning, intead
they yield a relatively precise and physically self-consistent calculation of
the outcome of cosmic evolution for a specific set of assumptions about star
formation and its regulation through feedback. The simulations show that it is
actually very hard to come up with a successful parameterization of the
feedback processes, arguably harder than it may seem based on the success of the
semi-analytic models. This clearly indicates that the effect of
  relevant feedback processes is still poorly understood in direct
  hydrodynamical simulations. However, one should also remember that the
  success of SAMs in reproducing luminosity function data is also the result of
  a tuning of the parameters entering the models.
 
\subsection{The properties of central galaxies}

To investigate the suitability of GIMIC, SAM and CLF galaxy populations for
fossil groups, we next consider properties of the brightest cluster galaxy
(BCG) and the second brightest galaxy. \cite{Yang2008}, based on the galaxy
groups extracted from SDSS DR4, produced two fitting formulae for the relation
between the luminosity of BCGs and the halo mass \citep[Eq. $6$
  in][]{Yang2008}, as well as for the relation of stellar mass of the most
massive galaxy (MMG) and the halo mass \citep[Eq. $7$ in][]{Yang2008}. The
halo masses in their group catalogues, and thus in the two fitting formulae,
are estimated though the ranking of the characteristic group luminosity $M_L$
and the ranking of the characteristic group stellar mass $M_S$ (see more
details in \citealt{Yang2007}). The two sets of halo masses agree reasonably
well with each other. Note however that the halo masses in their models are
obtained for the WMAP3 cosmology. For consistency, we here converted the halo
masses in their models to the ones corresponding to the MS using the abundance
matching method.  We use these converted fitting formulae as a reference
standard. The different halo masses of each model are normalized to the CLF
model, as described in Section~\ref{sec:3}. The selection of the central
galaxy data from our simulation is the same as in \cite{Yang2008}: either the
brightest cluster galaxy (BCG) or the most massive cluster galaxy (MCG) is
taken as the central galaxy.  For the ``isolated galaxies'', we apply a cut
with the magnitude limit of $\rmag < -16$.  To change to a common magnitude
system, we adopted for all models an $r$-band magnitude equal to $r = 4.76$
for the Sun in the AB system \citep{Blanton07}.

In the upper panels of Figure~\ref{fig:BCG}, we compare model predictions and
observational data for the relation between the $r$-band luminosity of the BCG
and the halo mass. In a similar way, in the lower panels we show the relation
between stellar mass of the MMG and halo mass. Different panels are for the
different model galaxy catalogues. In each panel, results are shown for the
regions at different overdensities.  The SAM models and the CLF model both
follow the observed relations reasonably well, while the GIMIC simulations
show substantial disagreements.  The tendency of GIMIC to overpredict the mass
of the MMG and the luminosity of the BCG begins at halo masses of $\sim
10^{13} \hMsun$. This overprediction occurs because GIMIC forms too many star
particles in the central galaxies at late times.
  
We now explore the gap between the first and second ranked galaxies.  In
Figure~\ref{fig:BCG_ratio}, we show the ratio between the brightest (massive)
and the second brightest galaxy versus the mass of the host halo, which is
directly related to the fossil characteristics of groups.  The cross and
diamond symbols included for comparison come from the SDSS \citep{Yang2008},
and correspond to two different treatments adopted in case the second
  galaxy in a group is not observed, because it is fainter than the SDSS
  magnitude limit of about $r\sim 17.7$.  In the first case, the second
  brightest (massive) galaxy is treated as having zero luminosity (mass), and
  is shown with crosses. In the second case, the second brightest (massive)
galaxy is treated as the magnitude limit of the SDSS survey (diamonds).  The
two SAM models and the CLF model all show a better fit to the observational
data than GIMIC at halo masses greater than $10^{13} \hMsun$.  In lower
density regions, the ratio in the CLF model and in the two SAMs also does not
show large deviations from the higher density regions, unlike GIMIC; but here
no tight observational constraints are available. As the observation prefered,
the more massive haloes tend to have smaller luminosity gaps. This is
supported by our catalogues, and also consistent with the findings of
\cite{Skibba07,Skibba11}.

\subsection{Halo occupation distribution}

\begin{figure}
\bc
\includegraphics[width=0.5\textwidth]{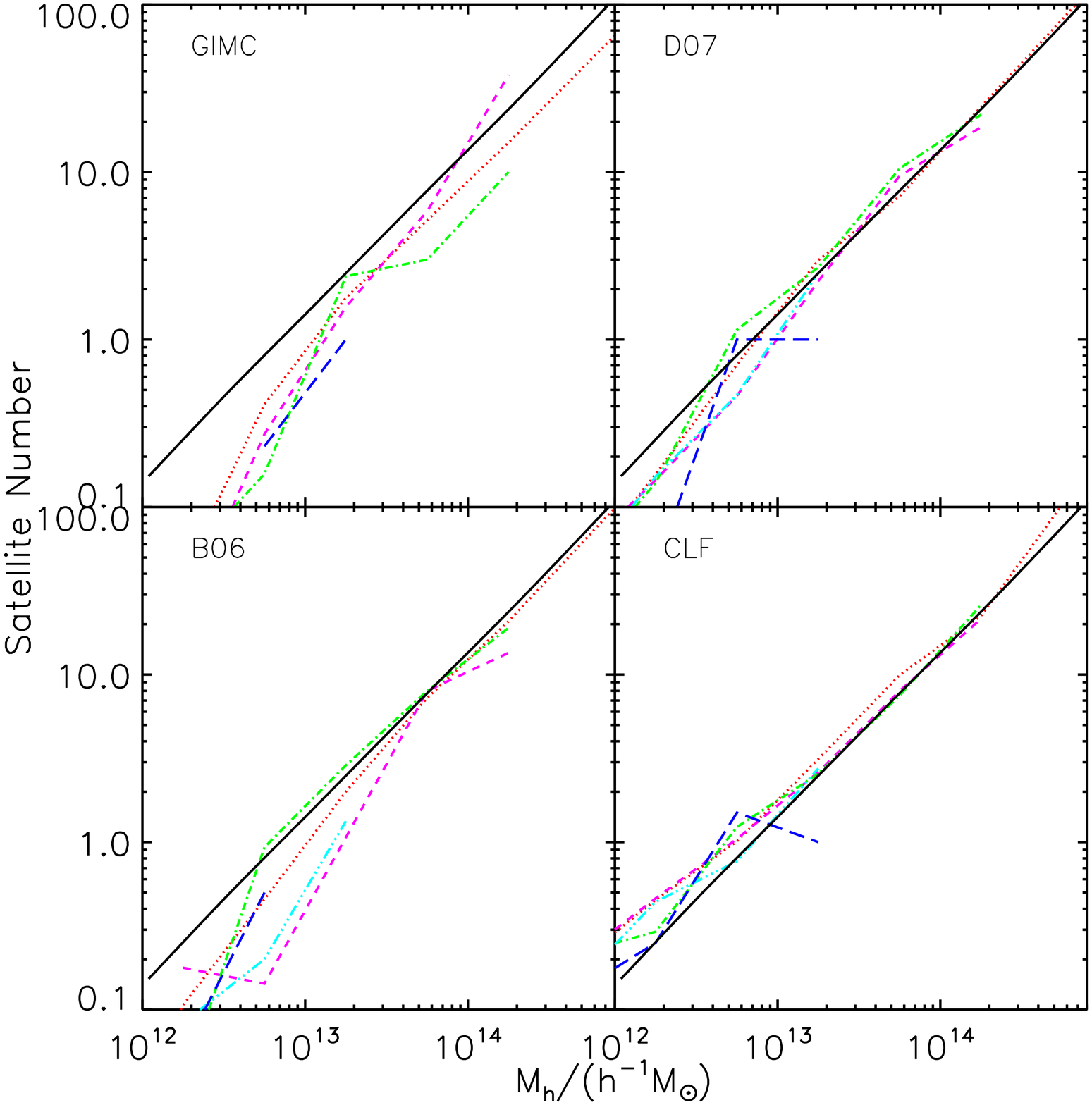}
\caption{ Mean satellite number as a function of halo mass. The colored lines
  encode the different density regions in the same way as in
  Fig. \ref{fig:BCG}, for an absolute magnitude limit $\rmag \leq -19.0$. The
  solid black line is the fitting result by \citet{Yang2008} for the same
  magnitude limit.
\label{fig:occupation}
}
\ec
\end{figure}

\begin{figure*}
\bc
\includegraphics[width=1.0\textwidth]{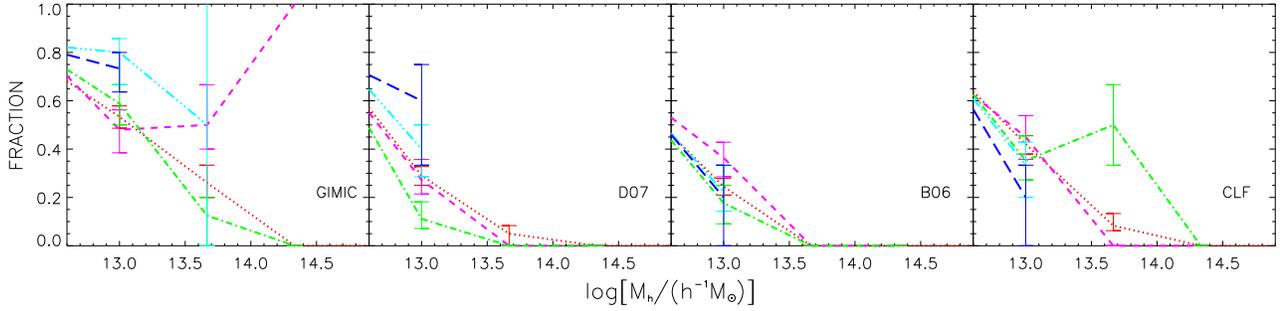}
\caption{ The optical fossil fraction as a function of halo mass for all four
  theoretical models we analyzed here. Each panel shows the results for one of
  the catalogues, as labeled. The lines with different colors and styles
  correspond to the different density regions, as in Fig.~\ref{fig:BCG}.
\label{fig:o_frac}
}
\ec
\end{figure*}

\begin{figure*}
\bc
\includegraphics[width=1.0\textwidth]{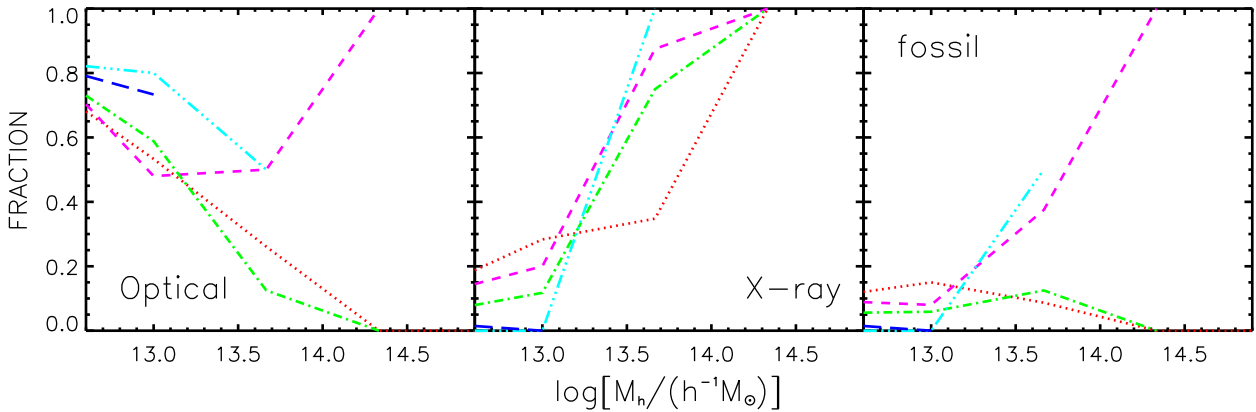}
\caption{ {\em Left panel:} The optical fossil fraction of GIMIC as a function
  of halos mass.  {\em Middle panel:} The fraction of GIMIC groups with X-ray
  luminosity $L_X$ larger than $2.13 \times 10^{42} \Xunit$.  {\em Right
    panel:} The overall fossil fraction when both selection criteria are
  combined.  In each panel, we show results for different overdensities using
  different colors and styles, based on the same key as in Fig.~\ref{fig:BCG}.
\label{fig:GIMIC_frac}
}
\ec
\end{figure*}

In Figure~\ref{fig:occupation}, we plot the satellite numbers versus halo
mass, with a specified magnitude limit, $\rmag \leq -19.0$, for all
models. The fitting formula of \citet{Yang2008} that we include for comparison
takes a power law form, $\langle N_s \rangle = (M_h/M_{s,0})^\gamma$, where
$\langle N_s \rangle$ is the mean number of satellite galaxies, $M_h$ is the
halo mass, and the two parameters $M_{s,0}$ and $\gamma$ have the best fitting
values of $10^{12.77}\,M_\odot$ and $1.06$, respectively.  After accounting
for the cosmology transformation from WMAP3 to Millennium for the fitting
formula, and a renormalization of the halo mass to CLF using halo abundance
matching method for all the models, the satellite numbers of all the models
are roughly consistent with the observed power--law fitting relation from
observational data, except at the low halo mass end.  Quite interestingly,
based on the results shown in Figure~\ref{fig:occupation}, we do not see any
evidence for environmental dependence of the halo occupation distribution in
any of the considered models.

\section{Results for fossil groups} \label{sec:5}

\subsection{Abundance of fossil groups in different environments} 

\begin{table*}
\begin{tabular}{ccccc}
\hline
Mass range ($\hMsun$) & Optical fossil fraction &  Fossil fraction & Reference
\\ 
\hline
$10^{13} - 10^{14}$ & $6.5\%$ &  - & \cite{Bosch} \\ 
$10^{12} - 10^{13}$ & $13.4\%$ &  -& \cite{Bosch} \\ 
$\sim 10^{13.5}$ & $11\%-20\%$ &  -& \cite{Yang2008} \\ 
$\sim 10^{13}$ & $18\%-60\%$ &  - & \cite{Yang2008} \\ 
$1-5 \times 10^{13}$ & $18\%$ &  -& \cite{Onghia2007} \\ 
$\geqslant 10^{13}$ & $8-10$ &  - & \cite{Sales} \\ 
$\sim 10^{13}-10^{15}$ & $13.3 \%$ & - & \cite{Dariush2007} \\
- & - & $7.2 \%$ & \cite{Dariush2007} \\ 
$\sim 10^{13}-10^{14}$ & $5-40 \%$ & - & \cite{M2006}\\ 
- & - & $8-20 \%$ & \cite{Jones2003}\\ 
$\sim 10^{14}$ & $33\%$ & - & \cite{Onghia2005,Jesper}\\
\hline
\end{tabular}
  \caption{Compilation of fossil fractions reported in the literature. In the
    first column we list the halo mass range for which the estimate applies,
    and where possible, we distinguish between optical and X-ray based fossil
    fractions. The last column gives the corresponding references, which here
    include both theoretical and observational works. \label{table:rf}}
\end{table*}

One of the most important parameters for fossil groups is their abundance.
Previous analyses have studied, for example, the 2-degree Field Galaxy
Redshift Survey \citep[e.g.][]{Bosch}, the SDSS-DR4 \citep{Yang2008},
semi-analytic galaxy catalogues \citep{Sales}, or cosmological N-body
simulations \citep{Onghia2007}. Since X-ray data has often been unavailable,
many works only considered the magnitude gap between the first and second
brightest galaxies, disregarding the requirement for a group to be also X-ray
luminous. We note that the predicted optical fossil fractions from these works
do not agree particularly well with each other, demonstrating that the
uncertainty in this quantity is still high (see Table~\ref{table:rf}).
Interestingly, several studies \citep{Onghia2005, M2006, Bosch, Yang2008,
  Dariush2007, Dariush2010} have consistently claimed a trend of an increasing
optical fossil fraction with decreasing halo mass.

We start by defining fossil groups based only on the magnitude gap in the
r-band, requiring it to be larger than 2 for an optical fossil group, which is
the commonly employed definition. In Figure~\ref{fig:o_frac}, we plot the
resulting average optical fossil fraction as a function of halo mass. A mass
trend is clearly present in all models. It is not surprising to see such a
declining trend in light of what we showed in Figure~\ref{fig:BCG_ratio}. The
strong disagreement in GIMIC (magenta line) is primarily related to the
sparseness of the data for the highest halo mass region, and the associated
statistical uncertainty. However, the optical fossil fractions in each model
are not fully consistent in detail. The error bars in Figure~\ref{fig:o_frac}
reflect the $1 \sigma$ scatter obtained from 200 bootstrap resamplings of
different groups. All models predict a very high optical fossil fraction at
the lower halo mass end. This means that large magnitude gaps become much more
common in low mass halos.  Taken into account the big error bars, we do not
find any clear evidence that environment has an effect on the optical fossil
fraction in any of the models.

As we discussed earlier, the hot gas profiles in the GIMIC halos can be used to
measure the X-ray luminosity $L_X$ of halos, providing us with a way to arrive
at higher specificity in the selection of fossil groups. We now define a group
to be fossil only if its X-ray luminosity $L_X$ is larger than $2.13 \times
10^{42} \Xunit$, in addition to showing the magnitude gap. This X-ray
  luminosity limit first adopted by \cite{Jones2003} is meant to exclude normal
  galaxies which are not elliptical galaxies at the centre of groups, and
  inevitably imposes a selection bias on the data. In
Figure~\ref{fig:GIMIC_frac}, we show the fossil fraction in GIMIC as a function
of halo mass based on this definition. In the middle panel of
figure~\ref{fig:GIMIC_frac}, we show the fraction of groups that have X-ray
luminosity $L_X$ above the threshold $2.13 \times 10^{42} \Xunit$. It is not
surprising to see that the fractions show an increasing trend with halo
mass. Although there are more optical fossil groups in low mass halos, the
fossil fraction is very small (less than $15 \%$) due to the limited X-ray
luminosity. In general, because of the additional X-ray selection,
  the halo mass dependence of the fossil groups seen in the optical band
  vanishes.

\begin{figure}
\bc
\includegraphics[width=0.5\textwidth]{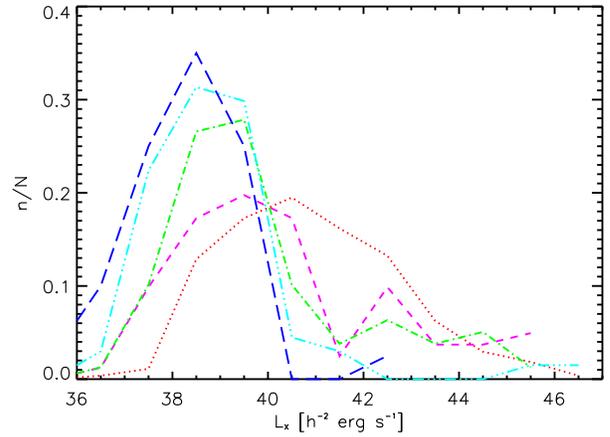}
\caption{The group abundance as a function of X-ray luminosity.  The quantity
  $n$ shown on the $y$-axis is the group number in each $L_X$ bin, while $N$
  is the total group number.  Differently colored and styles lines represent
  the different density regions, as in Fig.~\ref{fig:BCG}.
\label{fig:X-bin}
}
\ec
\end{figure}

\begin{figure*}
\bc
\includegraphics[width=1.0\textwidth]{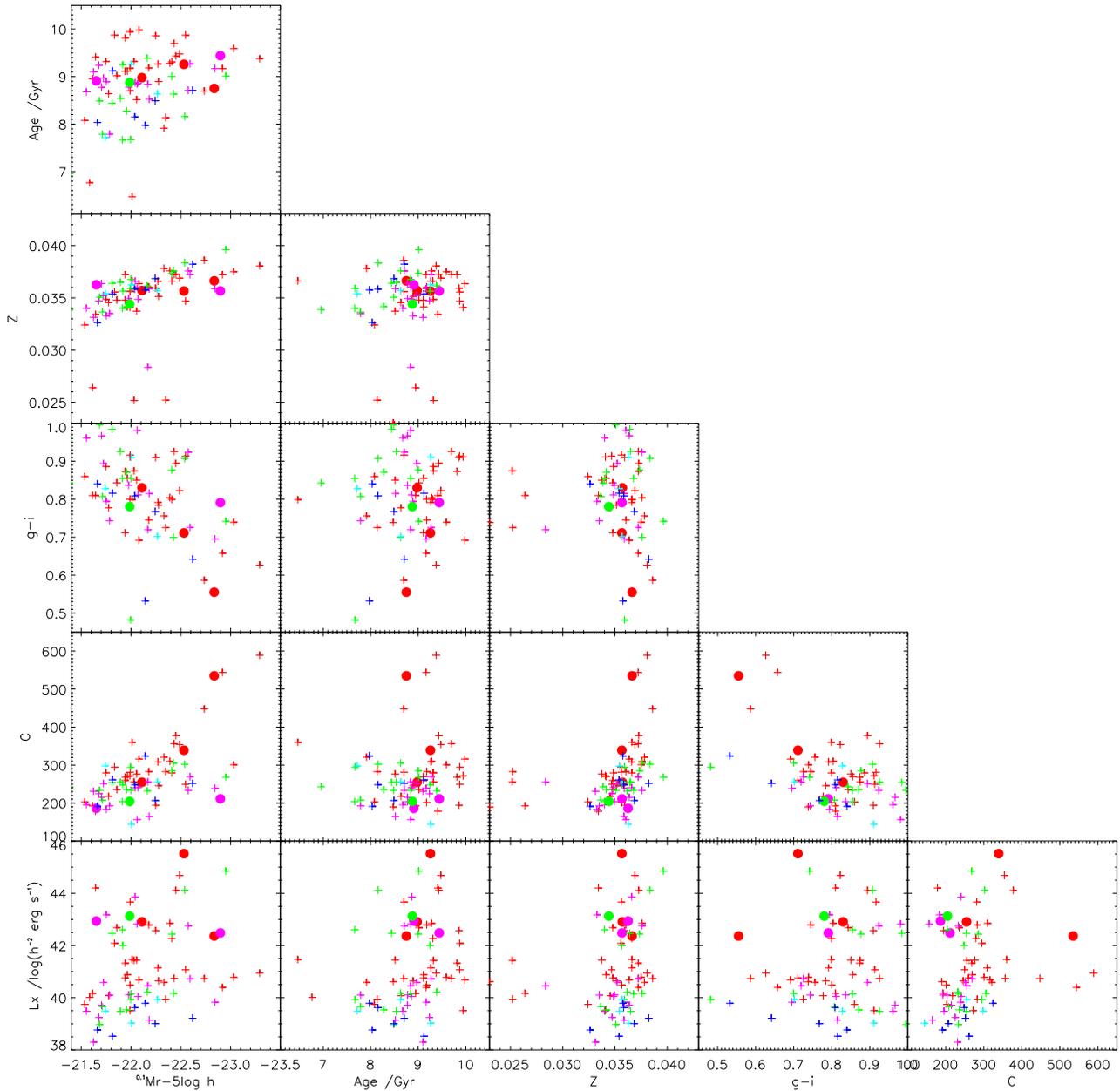}
\caption{Basic central galaxy properties  of fossil and non-fossil groups. The
  individual  panels give  the  relations between  six  quantities: the  X-ray
  luminosity $L_X$,  the $r$ band  magnitude $M_r$, the  mean age in  Gyrs, the
  mean metallicity  $Z$, the $(g-i)$ color, and  the galaxy  concentration $C$.
  Crosses  represent non-fossil  groups, while  solid circles  are  for fossil
  groups. The  different colors refer to  the different density  regions as in
  our previous figures.
\label{fig:all}
}
\ec
\end{figure*}

Figure~\ref{fig:GIMIC_frac} also reveals that X-ray bright groups in low
density regions are very rare, which appears consistent with the dramatic drop
found for low mass halos by \citet[][see their Figure 5]{Dariush2007}. To
investigate this reduction of the X-ray fossil fraction in two lower density
regions, we further plot the abundance of the groups in $L_X$-bins for the
halo mass range $10^{12} - 10^{15} \hMsun$ in Figure~\ref{fig:X-bin}. Most of
the X-ray luminous groups in the two lower density regions are located around
$L_X \sim 10^{38}\Xunit$. There are however very rare galaxy groups with $L_X
> 2.13\times 10^{42}\Xunit$ in the two lowest density regions. The abundance
of systems with $L_X > 2.13\times 10^{42}\Xunit$ within the average density
and high density regions are much higher than in the low density regions.  The
main reason for this behaviour is that structures in low-density regions are
generally characterized by shallower potential wells. Therefore, the gas
reaches relatively lower densities in the central part of these structures,
thus explaining the correspondingly lower X--ray luminosity.

\subsection{The properties of fossil groups}

We now focus on the halo mass range $9.0 \times 10^{12} \sim 4.0 \times
10^{13} h^{-1}\Msun$, which is likely the most reliable in GIMIC.  In
Figure~\ref{fig:all}, we summarize our measurements for some of the basic
properties of the central galaxies of fossil groups and non-fossil groups,
such as age, metallicity, colors or concentration, and we discuss them in the
following.
 
\citet{Jones2003} found that fossil groups are more X-ray luminous for a given
optical luminosity than non-fossil groups. Using {\it Chandra} X-ray
observations of five fossil groups, supplemented by additional systems from
the literature, \citet{Habib} confirmed their result of a $L_X$-$L_R$
relation.  However, a recent work by \citet{Alexey} reached a different
conclusion.  They claim that the comparisons in \cite{Jones2003} and
\cite{Habib} were affected by systematic errors due to a non-uniform
selection, because the data for fossil groups and other systems were obtained
separately. Using instead the data from a uniform sample, \cite{Alexey}
concluded that fossil groups are not systematically brighter in X-rays than
other groups at the same optical luminosity.  Our results in
Fig.~\ref{fig:all} for the $M_r$-$L_X$ relation support the conclusion by
\cite{Alexey}. In the region $L_X > 2.13 \times 10^{42}\Xunit$, we do not see
a systematic increase of the X-ray luminosity for fossil groups at a given
optical magnitude.

To quantify the age of our simulated galaxies, we directly use the mean age of
the stellar particles. Our results for this analysis are shown in the third
row of Figure~\ref{fig:all}.  Fossil groups in these plots typically have a
mean age around $9\, {\rm Gyr}$, but this age is not significantly enlarged
compared to the age of non-fossil groups. This suggests that there may be
little justification in the end to call the magnitude-gap selected sample of
groups ``fossil'' -- they do not appear to have different ages than normal
groups.

Similarly, the mean galaxy metallicities for fossil and non-fossil groups
reported in the fourth row of Figure~\ref{fig:all} show no apparent
differences.  We have also used our spectral synthesis code to calculate
colors for the simulated galaxies. The $g-i$ color is shown in the fifth row
of Figure~\ref{fig:all}. We here find values that are systematically lower
than observational results. That is because we have not introduced any dust
attenuation model in our synthesis code, and also because the simulation tends
to have too much recent star formation in central galaxies. This is also the
reason why the colors of our simulated fossil groups spread over a large range
from $0.5$ to $1.0$ in the plots. Many of them are actually too blue due to
recent star formation to represent bona-fide early type galaxies, even though
their large stellar mass otherwise suggests that they are.

Finally, the last row in Figure~\ref{fig:all} shows the concentration $C$ of
the central galaxies.  We here used the maximum circular velocity $V_{\rm
  max}$ of the central galaxy's halo and the radius $r_{\rm max}$ at which
this velocity is attained to estimate halo concentrations \citep{Aquarius}.
Note that this procedure assumes that the dark matter halo is not
substantially modified by the settling of the baryons in the center, and that
the rotation curve is dominated by dark matter at its maximum; both of these
assumptions may actually break down if the central galaxies are very luminous
and compact. Regardless of this complication, we may still go ahead and
compare the concentration estimates obtained in this way among the different
regions, as done in Fig.~\ref{fig:all}.  We see that although there is one
fossil galaxy in the $+2 \sigma$ region that shows an unusually high
concentration value, most fossil groups are well intermixed with the
distribution of normal groups, providing no evidence for a significant
difference in the concentrations.

\begin{figure}
\bc
\includegraphics[width=0.5\textwidth]{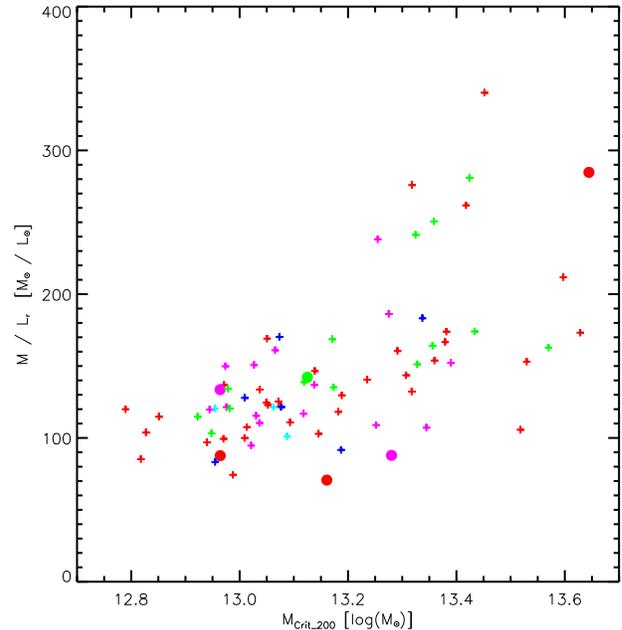}
\caption{ Mass-to-light ratio in the GIMIC simulations as a function of galaxy
  stellar mass.  The crosses are central galaxy's mass-to-light ratio from
  non-fossil groups, while the solid circles show fossil groups. The different
  colors encode the different density regions, as in the previous figures.
\label{fig:ml}
}
\ec
\end{figure}

We have also investigated the mass-to-light ratios of the central galaxies in
the selected GIMIC data.  Previous works \citep[e.g.][]{Habib, Vikhlinin,
  Yoshioka} have found conflicting results for the mass-to-light ratio of
fossil groups. Using $M/L_R$ \citet{Vikhlinin} reported a mass-to-light ratio
$\sim 250-450 M_{\odot}/L_{\odot}$, and using $M_{\rm crit200}/L_B$
\cite{Yoshioka} found $\sim 100-1000 M_{\odot}/L_{\odot}$.  Both of these
studies claimed a very high mass-to-light ratio for the X-ray overluminous
elliptical galaxies, while using $M_{\rm vir}/L_B$, \cite{Habib} reported that
the mass-to-light ratio of fossils tends to be at the upper envelope of the
values seen in normal groups and clusters. In Figure~\ref{fig:ml}, we show our
measurements for the GIMIC simulation. We use $M_{\rm crit200}$ as the group
mass to calculate the mass-to-light ratio. Here we also find that the
mass-to-light ratio for fossil groups is consistent with that found for
non-fossil groups.
 
\begin{figure}
\bc
\includegraphics[width=0.5\textwidth]{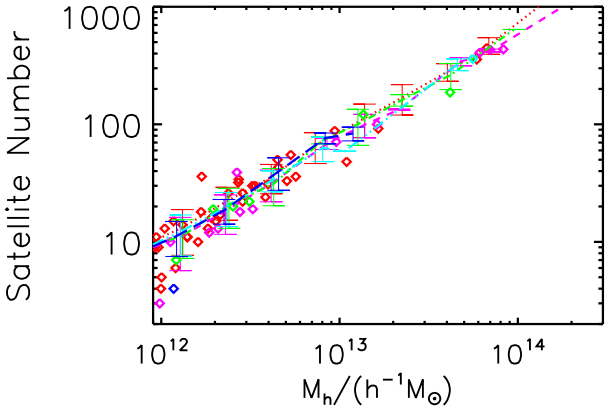}
\caption{ The number distribution of all satellite galaxies in the
  GIMIC simulations. The different lines are the mean satellite number for non
  fossil groups, while the diamonds show results for fossil groups. The
  different colors correspond to different density regions, as in the previous
  figures. The error bars indicate the r.m.s. scatter within each mass bin.
\label{fig:cksn}
}
\ec
\end{figure}

Based on our results, there is hence little tangible evidence for differences
in the properties of fossil and non-fossil groups. In \cite{Barbera}, they
compared both structural and stellar population properties of fossil and
non-fossil group galaxies, which is also shown consistent results. This calls
into question whether the fossil groups are indeed a special class of objects
that is characterized by an unusual formation history. Instead, our results
appear more consistent with an interpretation of fossil groups as a common
phase in galaxy evolution during which they lie in the tail of the distribution
of a certain group properties, in this case the magnitude difference between
first and second ranked galaxies. This result is in line with findings
  obtained by \cite{ Dariush2010}, who also used data from the Millennium
  simulation. These authors found that, no matter at what redshift the fossil
  groups are selected, after $\sim4$Gyr more than $\sim 90$ percent of them
  change their status and become non-fossil according to the magnitude gap
  criterion. This behavior, obtained from the galaxy formation histories,
  directly supports our conclusion here. 
  As a further check, we show in Figure~\ref{fig:cksn}
  the number of satellites for fossil and non fossil groups.  Here we use all
  the satellite galaxies in the GIMIC simulations. The
  satellite number for fossil groups is not different from that obtained
  considering non fossil groups, which confirms that the fossil groups do not
  have unusual formation histories. Nevertheless, we stress that although the
  fossil groups defined according to the magnitude gap are statistically
  similar to non fossil groups, there may be rare situations in which the
  fossil groups indeed have unusual formation histories (the few outliers
  visible in Figure~\ref{fig:cksn}).
 
\section{Discussion and conclusions} \label{sec:6}

In this study, we have analyzed fossil groups in the hydrodynamic simulation
data of GIMIC combined with CLF catalogues and two semi-analytic galaxy
formation models. Our primary goal has been to shed light on the question of
how well models of galaxy formation reproduce the observational properties of
fossil groups, and whether there is any evidence that this class of objects is
characterized by a special formation history.  To this end we have first
investigated the luminosity functions of our theoretical models, and in
particular the abundance of brightest and most massive cluster galaxies. We
then considered the ratio of luminosity or stellar mass between the first and
second ranked cluster galaxies, and examined the satellite abundance as a
function of host halo mass. In all of our measurements, we have also checked
whether there is any environmental difference between the five regions
simulated by GIMIC, which have substantially different mean densities.

For defining ``fossil groups'', we have mostly employed the luminosity gap
between the two first ranked galaxies in a group, similar to what has been
done in many previous works. However, by also measuring the X-ray luminosity
of halos in GIMIC, we could more sharply define fossil groups as those which
also have high X-ray luminosity, reflecting more closely the originally used
observational definition. Since X-ray emission is unavailable in the CLF and
the SAMs, we have however applied this selection only in parts of our
analysis. Finally, we have investigated some basic properties of fossil and
non-fossil groups in order to see in which properties they differ most
significantly. For this, we restricted our halo sample to the mass range $9.0
\times 10^{12} \sim 4.0 \times 10^{13}\hMsun$, which should be the most
reliable in our hydrodynamical simulations.

\noindent Our primary conclusions can be summarized as follows.
\begin{itemize}

\item The optical fossil fraction in all of our theoretical models declines
  with increasing halo mass.  However, the models do not agree in detail at a
  specific halo mass. And we did not find clear evidence that the optical
  fossil fraction has environmental effects in all the theoretical galaxy
  formation models.

\item After applying the bright X-ray luminosity limit in the
    selection of fossil groups, the total fraction of fossil groups
    decreases significantly in small halos. As a consequence, the halo
    mass dependence of the fossil groups seen in optical vanishes.

\item From the $L_X$ distribution of groups identified in the GIMIC
  simulations, as shown in Figure~\ref{fig:X-bin}, we have found evidence for a
  clear environmental effect: halos found in lower-density regions have
    relatively lower X-ray luminosity. Their median X-ray luminosity can vary
    by two orders of magnifude, from $\sim 10^{38.5}\Xunit$ ($-2 \sigma$
    overdensity region in GIMIC) to $\sim 10^{40.5}\Xunit$ ($2 \sigma$
    overdensity region in GIMIC).

\item When we investigated the properties of central galaxies in fossil and
  non-fossil groups, we have found no differences in the magnitude, X-ray
  luminosity, age, metallicity, concentration, color or mass-to-light
  ratio. In particular, the mean ages of the central galaxies in fossil groups
  are not really 'fossil' in the sense of the word. This casts doubts about
  whether fossil groups are a useful concept to identify a particular class of
  early type galaxies.

\item In addition, we have checked that the satellite galaxies number 
  distribution of fossil and non-fossil groups. There is not significantly 
  different difference between these two populations as well.

\end{itemize}

Our results are consistent with the findings of \citet{Beckmann}, who studied a
concordance $\Lambda$CDM cosmological simulation and pointed out that many
groups will go through an `optical fossil phase' that is typically ended by
renewed infall from the environment. In this picture, fossil groups are simply
groups that temporarily are in a `fossil phase', such that a significant
difference in the central galaxy properties can not be expected. It is
  consistent with the analysis carried out by \cite{Dariush2010}, using also
  Millennium data, who found that about $90\%$ fossil groups will
  become non-fossils after $\sim4$Gyr. The other important result of our
analysis is the absence of strong environmental effects at the group scale, at
least at the level that could be probed with the limited sample sizes that we
had available in the GIMIC regions. It is well possible that there are still
weak trends with large-scale overdensity, but uncovering those reliably will
require larger simulation volumes.
 
\section*{Acknowledgements} 

The authors would like to thank Lei Liu for the help and discussion of SAM
data, St\'{e}phane Charlot for the help of SSP templates of \cite{BC03}, and
the anonymous referee for constructive comments that greatly improved the
presentation of this paper. We also thank Youcai Zhang, Jing Wang, Shiyin
Shen, Till Sawala, Klaus Dolag, Pierluigi Monaco for valuable discussions. The
semi-analytic galaxy catalogue is publicly available at
http://www.g-vo.org/MyMillennium3/

Weiguang Cui acknowledges a fellowship from the European Commission's
Framework Programme 7, through the Marie Curie Initial Training Network
CosmoComp (PITN-GA-2009-238356). GDL acknowledges financial support from the
European Research Council under the European Community's Seventh Framework
Programme (FP7/2007-2013)/ERC grant agreement n. 202781. This work is partly
supported by 973 Program (No. 2007CB815402), the CAS Knowledge Innovation
Program (Grant No.  KJCX2-YW-T05), grants from NSFC (Nos. 10821302, 10925314),
by a PRIN09-INAF grant and by the INFN-PD51 grant.

\bibliography{paper_arxiv}
\bibliographystyle{mn2e.bst}

\end{document}